\documentstyle[aps,pra,subfigure,epsfig,amsmath,amssymb]{revtex}

\begin{document}
\draft
\title{Temperature dependence of density profiles for a cloud 
of non-interacting fermions moving inside a harmonic trap in one dimension}
\author{Z. Akdeniz$^{1,2}$, P. Vignolo$^1$, A. Minguzzi$^1$ and M. P. Tosi$^1$}
\address{(1) Istituto Nazionale di Fisica della Materia and Classe di Scienze,
Scuola Normale Superiore,
Piazza dei Cavalieri 7, I-56126 Pisa, Italy\\
(2) Department of Physics, University of Istanbul, Istanbul, Turkey}
\maketitle

\begin{abstract}
We extend to finite temperature a Green's function method that was previously 
proposed to evaluate ground-state properties of mesoscopic clouds of 
non-interacting fermions 
moving under harmonic confinement in one dimension. 
By calculations of the particle and 
kinetic energy density profiles we illustrate the role 
of thermal excitations in smoothing out the 
quantum shell structure of the cloud and in spreading 
the particle spill-out from quantum tunnel 
at the edges. We also discuss the approach of the exact 
density profiles to the predictions of a 
semiclassical model often used in the theory of confined 
atomic gases at finite temperature.
\end{abstract}

\vspace{0.2cm}

\section{Introduction} 
The achievement of Bose-Einstein condensation in ultracold gases of bosonic 
atoms has given great impulse to the study of dilute quantum gases 
inside magnetic or optical traps \cite{ref1}. 
Similar cooling and trapping techniques are being used to drive gases 
of fermionic alkali atoms 
into the quantum degeneracy regime \cite{ref2}. It is also possible to 
experimentally generate and study 
strongly anisotropic atomic fluids effectively approaching dimensionality 
$D = 2$ or $D = 1$, at very low temperature and with very high purity, 
inside magnetic traps where the transverse and the 
longitudinal confinement are vastly different \cite{ref3}.
	
In the relevant conditions of temperature and dilution the atomic 
interactions become negligible in a gas of fermions placed in a single 
Zeeman sublevel inside a magnetic trap \cite{ref4}. 
One can obtain in this way a close laboratory realization of an 
inhomogeneous, non-interacting Fermi gas, which has special significance 
in regard to the foundations of density functional 
theory and to tests of the Thomas-Fermi approximation \cite{ref5}. 
In the quasi-onedimensional (1D) 
case the ground state and the excitation spectrum of such an ideal 
gas of spin-polarized (or "spinless") fermions can be mapped into those
 of a gas of hard-core impenetrable bosons \cite{ref6}. 
The latter is known as the Tonks gas, from the early work of 
Tonks \cite{ref7} on the equation of state of hard-object adsorbates. 
Advances in atom waveguide technology, with potential applications 
to atom interferometry and integrated atom optics, especially motivate 
theoretical studies of dilute gases in a regime where the dynamics 
becomes essentially 1D \cite{ref8}.
	
The wave functions of a cloud of spinless fermions under harmonic 
confinement are, of course, Slater determinants of harmonic-oscillator 
single-particle orbitals. Their representation in terms of Hermite 
polynomials has, however, limited usefulness for numerical calculations on 
mesoscopic clouds \cite{ref9}. Brack and van Zyl \cite{ref10} have 
developed a more powerful method for non-interacting fermions occupying 
a set of closed shells under isotropic harmonic confinement 
in D dimensions, leading to analytical expressions for the particle and 
kinetic energy densities at temperature $T = 0$ in terms of Laguerre 
polynomials. These expressions are especially useful for 
numerical calculations on isotropic systems in $D = 2$ and $D = 3$. 
A Green's function method, which altogether avoids the use of wave 
functions in favor of the matrix elements of the position 
and momentum operators, has been developed for similar purposes \cite{ref11} 
and extended to calculations of the pair distribution function in the 
ground state for the 1D ideal Fermi gas \cite{ref12}. 
This method has also been extended to the ground state of ideal Fermi 
gases under harmonic confinement of arbitrary anisotropy in higher 
dimensionalities \cite{ref13}.
	
The purpose of the present paper is to extend the Green's function 
method to an inhomogeneous 1D cloud of non-interacting fermions at 
finite temperature and to illustrate its usefulness by numerical 
calculations of the particle and kinetic energy density profiles as 
functions of temperature in the case of harmonic confinement. 
Analytical expressions have been derived by Wang\cite{ref14} for 
the same system at high temperature, where the chemical potential is 
lower than the single-particle ground-state energy. The case of 
non-interacting fermions under 3D harmonic confinement has been treated 
by Schneider and Wallis \cite{ref15} through the use of 
Laguerre polynomials. The emphasis of our numerical calculations 
will be to illustrate how the characteristic quantum features of the 
fermion cloud, {\it i.e.} its shell structure and the spill-out of 
particles at the boundaries beyond the Thomas-Fermi radius, 
evolve with increasing temperature as a semiclassical regime is being 
approached.

	The paper is organized as follows. 
Section II reports some essential definitions, starting 
from the one-body density matrix at $T\neq 0$, and gives the expressions 
of the density profiles in the semiclassical regime. 
Sections III and IV present the essential details of our method and our 
numerical results, respectively. 
A brief summary concludes the paper in Section V. 
Although we refer through the paper to the results as being appropriate 
to spinless fermions in 1D, they are equally valid for the Tonks gas 
of impenetrable bosons.

\section{ESSENTIAL DEFINITIONS AND THE SEMICLASSICAL LIMIT}
The generalized grand-canonical density matrix for fermions~\cite{ref16}
can be written as

\begin{equation}
D(x_1,x;\beta,\mu)=\sum_{i=1}^\infty\frac{1}{1+\exp{[\beta(E_i-\mu)]}}
\psi_i^*(x_1)e^{i\hat{p}(x-x_1)}\psi_i(x_1).
\label{rho_def}
\end{equation}
Here $\beta=1/k_BT$ and $\mu$ is the chemical potential, while 
$\psi_i$ and $E_i$ are the single-particle orbitals and
the corresponding energy eigenvalues.
The zero temperature limit
of Eq.~(\ref{rho_def}) leads to the Dirac density matrix for $N$ ideal 
spinless fermions,
\begin{equation}
\rho(x_1,x)=\sum_{i=1}^N \psi_i^*(x_1)e^{i\hat{p}(x-x_1)}\psi_i(x_1).
\end{equation}

The particle density profile $n(x)$ of the gas at temperature $T$ and 
chemical potential $\mu$ is the zero-order moment of the matrix $D(x,x_1)$,
\begin{eqnarray}
n(x)&=&D(x_1,x;\beta,\mu)|_{x_1=x}\nonumber\\&=&
\sum_{i=1}^\infty\frac{1}{1+\exp{[\beta(E_i-\mu)]}}\langle \psi_i\,|
\delta(x-\hat{x})|\,\psi_i\rangle.
\label{dens}
\end{eqnarray}
The kinetic pressure $P(x)$ is given by a specific second-order moment 
of $D(x,x_1)$ \cite{ref17},
\begin{eqnarray}
P(x)&=&-\frac{\hbar^2}{m}\frac{\partial^2}{\partial x_1^2}
D(x_1,x;\beta,\mu)|_{x_1=x}\nonumber\\&=&
\frac{1}{2m}\sum_{i=1}^\infty\frac{1}{1+\exp{[\beta(E_i-\mu)]}}
\langle \psi_i\,|\hat{p}^2\delta(x-\hat{x})+\delta(x-\hat{x})\hat{p}^2|
\,\psi_i\rangle.
\label{press}
\end{eqnarray}
The kinetic pressure $P(x)$ is twice the kinetic energy density of the 
fermion cloud.
In these equations $\hat{p}$ and $\hat{x}$  are 
the momentum and position operators, and the chemical potential
$\mu$ is determined from the condition
\begin{equation}
\int n(x;\beta,\mu)dx=N
\label{norm}
\end{equation}
where $N$ is the average number of particles.

In the semiclassical regime the particle density and the kinetic pressure 
of an ideal Fermi gas confined by a 1D potential $V(x)$ can be calculated 
by treating the energy levels as a continuum. The appropriate condition 
of validity is that the level spacing $\Delta E$ is sufficiently 
smaller than the thermal energy $k_BT$. The appropriate expressions in 
the grand-canonical ensemble are
\begin{eqnarray}
n_{sc}(x)&=&
\int dE \int\frac{dp}{2\pi\hbar}\delta\left(E-\frac{p^2}{2m}-V(x)\right)
\frac{1}{1+\exp[\beta(E-\mu)]}
\nonumber\\
&=& 
\int\frac{dp}{2\pi\hbar}\left\{\exp\left[\beta\left(
\frac{p^2}{2m}+V(x)-\mu\right)\right]+1\right\}^{-1}
\label{semi-class_nx}
\end{eqnarray}
and
\begin{equation}
P_{sc}(x)=
\int\frac{dp}{2\pi\hbar}\,\frac{p^2}{m}
\left\{\exp\left[\beta\left(
\frac{p^2}{2m}+V(x)-\mu\right)\right]+1\right\}^{-1}.
\label{semi-class_px}
\end{equation}
The chemical potential is again determined by normalization to the 
average number of fermions.

\section{The Green's function method}
Equations~(\ref{dens}) and (\ref{press}) can be rewritten in terms
of the Green's function 
$\hat{G}(x)=\lim_{\varepsilon\rightarrow 0^+}(x-\hat{x}+i\varepsilon)^{-1}$ 
in coordinate space,
\begin{eqnarray}
n(x)&=&-\frac{1}{\pi}\lim_{\varepsilon\rightarrow 0^+}
{\rm Im}\sum_{i=1}^\infty\frac{1}{1+\exp{[\beta(E_i-\mu)]}}\langle \psi_i\,|
\hat{G}(x)|\,\psi_i\rangle\nonumber\\&=&
-\frac{1}{\pi}\lim_{\varepsilon\rightarrow 0^+}
{\rm Im}{\rm Tr}\,({\cal T}\cdot\hat{G}(x))
\label{eq_nx}
\end{eqnarray}
and
\begin{eqnarray}
P(x)&=&-\frac{1}{\pi}\lim_{\varepsilon\rightarrow 0^+}
{\rm Im}\sum_{i=1}^\infty\frac{1}{1+\exp{[\beta(E_i-\mu)]}}\langle \psi_i\,|
\frac{\hat{p}^2}{m}\hat{G}(x)|\,\psi_i\rangle\nonumber\\&=&
-\frac{1}{\pi m}\lim_{\varepsilon\rightarrow 0^+}
{\rm Im}{\rm Tr}\,({\cal T}\cdot\hat{p}^2\cdot\hat{G}(x)).
\label{eq_px}
\end{eqnarray}
Here we have used a matrix formalism, by introducing the 
{\it temperature matrix} ${\cal T}$  whose 
diagonal elements $[{\cal T}]_{i,i}=1/\{1+\exp{[\beta(E_i-\mu)]}\}^{-1}$ 
contain the statistical Fermi factors, while the off-diagonal elements 
$[{\cal T}]_{i,j}$ with $i\neq j$ are null.

The evaluation of the expressions in Eqs. (\ref{eq_nx}) and (\ref{eq_px}) 
is carried out by an immediate extension of the procedure developed 
in \cite{ref11} for the case of zero temperature. The trace of a 
generic matrix $A$ is connected to the elements of the inverse matrix $A^{-1}$
 by the relation
\begin{equation}
{\rm Tr}A=\frac{\partial}{\partial\lambda}\left.
[\ln\det(A^{-1}+\lambda{\Bbb{I}})]\right|
_{\lambda=0}.
\end{equation}
We get from Eqs. (\ref{eq_nx}) and (\ref{eq_px})
\begin{equation}
n(x)=-\frac{1}{\pi}\lim_{\varepsilon\rightarrow 0^+}
{\rm Im}\frac{\partial}{\partial\lambda}\left.[\ln\det(x-\hat{x}
+\lambda {\cal T}+i\varepsilon)]\right|_{\lambda=0}
\label{eq2_nx}
\end{equation}
and
\begin{equation}
P(x)=-\frac{1}{\pi m}\lim_{\varepsilon\rightarrow 0^+}
{\rm Im}\frac{\partial}{\partial\lambda}\left.
[\ln\det(x-\hat{x}+i\varepsilon+\lambda {\cal T}\cdot\hat{p}^2)]\right|_{\lambda=0}.
\label{eq2_px}
\end{equation}
In the specific case of harmonic confinement  
we make use of the representation of
the position and the momentum operators 
in the basis of the eigenstates of the harmonic oscillator with
energy $E_n=(n-1/2)\hbar\omega$: that is,
$\hat{x}=(a+a^{\dag})/\sqrt 2$ and
$\hat{p} =i(a^{\dag}-a)/\sqrt 2$
with $a\,|\,\psi_n\rangle=\sqrt{n-1}\,|\,\psi_{n-1}\rangle$ and
$a^{\dag}\,|\,\psi_n\rangle=\sqrt{n}\,|\,\psi_{n+1}\rangle$.
The tridiagonal form of the matrices representing $\hat{x}$ and $\hat{p}$ 
allows us to express the determinants in Eqs. (\ref{eq2_nx}) 
and  (\ref{eq2_px}) by a recursive algorithm~\cite{ref11,ref17} 
as products of an infinite number of matrices having dimension
$1\times 1$ for the particle density
and $2\times 2$ for the kinetic pressure.
If we write the matrix 
${\cal K}^{\nu}=\hat{x}-\lambda{\cal T}\cdot\hat{p}^{\nu}$ in the 
tridiagonal form
\begin{equation}
{\cal K}^{\nu}=
\left(
\begin{matrix}
{\cal   A}_1 & {\cal B}_{1,2} &     & \\
{\cal  B}_{2,1} & {\cal A}_2 &{\cal  B}_{2,3} & \\
            & \ddots & \ddots & \ddots \\
\end{matrix}
\right)
\label{tridiag}
\end{equation}
where the indices refer to blocs of dimension $1\times 1$ if $\nu=0$
and $2\times 2$ if $\nu=2$, we obtain
\begin{equation}
\det(x-{\cal K}^{\nu}+i\varepsilon)=
\prod_{j=1}^{\infty}\det(x-\tilde{\cal A}_j+i\varepsilon).
\label{INFM}
\end{equation}
Here $\tilde{\cal A}_1={\cal A}_1$ while the renormalized blocs 
$\tilde{\cal A}_j$ for $j>1$  are given by the recursive formula
\begin{equation}
\tilde{\cal A}_j={\cal A}_j+{\cal B}_{j,j-1}
(x-\tilde{\cal A}_{j-1}+i\varepsilon)^{-1}{\cal B}_{j-1,j}.
\label{ric}
\end{equation}
As in the calculation of density profiles at $T = 0$ \cite{ref11} 
the evaluation of the determinant in Eq. 
(\ref{INFM}) converges quite rapidly, yielding accurate results 
for mesoscopic fermion clouds with a moderate amount of numerical effort.

\section{Numerical results}
We report in this section some numerical results for the particle density 
and the kinetic pressure that we have obtained by the Green's 
function method. Our main purpose is to illustrate 
the approach of the profiles to their semiclassical expressions 
and we consider for clarity clouds 
containing rather small average numbers of particles, {\it i.e.} 
$N = 4$ and $N = 20$. 
Some sharpening of the structures in the profiles should be expected
in a canonical ensemble viewpoint, where
fluctuations in the particle number are suppressed.
	
In the calculations reported below we have approximated the determinant 
in Eq. (\ref{INFM}) by the product of its first $M$ terms with $M = 10^7$ 
as an upper bound. This choice allows for up to 
$10^7$ thermally excited states and the corresponding value of the 
spectral resolution parameter is chosen as $\varepsilon = 10^{-3}$. 
These two choices evidently limit to some minor extent the accuracy of our 
numerical results relatively to the formally exact theory given in Sect. III. 
Of course, for such small numbers of particles one may also approach the 
same calculations by using directly the expression of the harmonic-oscillator 
orbitals in Eq. (\ref{rho_def}). However, the routines that are available 
to evaluate Hermite polynomials also involve numerical approximations 
which are increasingly severe for polynomials of high degree, 
as needed to describe the gas at finite temperature.

The first step in our calculations is to evaluate the chemical potential 
at fixed $N$ by means of a self-consistent solution of Eqs. (\ref{norm}) 
and (\ref{semi-class_nx})~\cite{ref18}. The resulting semiclassical value of $\mu$ 
is used in Eqs. (\ref{eq2_nx}) and (\ref{eq2_px}) for the evaluation of 
the particle density and kinetic pressure profiles. The 
procedure is justified {\it a posteriori} by the fact that integration of 
the spatial density $n(x)$ obtained at this stage reproduces the correct 
average number 
$N$ of fermions.
	
In Figures 1 and 2 we report the particle density profiles at various 
temperatures for $N = 4$ and $N = 20$, respectively. 
The corresponding profiles of kinetic pressure are given in Figures 3 
and 4. In each Figure the left-hand panel shows how the quantum effects, 
which consist of oscillations from shell structure in the profiles and 
in particle spill-out and negative kinetic pressure at the edges, 
are washed away by thermal excitations. The right-hand panels show 
instead how the profiles approach the semiclassical regime: 
the latter already holds to a good approximation when $k_BT\simeq\hbar\omega$,
although some trace of a negative kinetic pressure in the spill-out region
remains in the ``exact'' profile at this temperature.

\section{CONCLUSIONS}
In conclusion, in this paper we have extended to finite temperature the 
Green's function method that was previously developed for the formally 
exact evaluation of ground-state properties of an inhomogeneous ideal 
Fermi gas in 1D. The method dispenses with knowledge 
of the single-particle orbitals in favour of the single-particle energy 
levels and of the matrix elements of the position and momentum operators. 
Our numerical applications to clouds containing small numbers of particles 
under harmonic confinement have shown that the commonly used semiclassical 
approach to a thermally excited fermion cloud yields (i) accurate 
values of the chemical potential except at the lowest temperatures, 
and (ii) a reliable account of particle 
and kinetic energy density profiles once their main quantum features 
become essentialy washed out at $k_BT\simeq\hbar\omega$.
	
Two further comments are in order. Firstly, the profiles shown in 
Figures 1 and 2 also describe (aside from an immediate rescaling of the 
units used on the axes) the momentum distribution of the fermion cloud 
under harmonic confinement. Secondly, the same profiles also 
refer to the Tonks gas of impenetrable bosons in 1D, as was explicitly 
shown by Yang and Yang \cite{ref19}.

\acknowledgments
This work was partially supported by INFM through the PRA-Photonmatter program.

\begin{figure}
\centering{
\epsfig{file=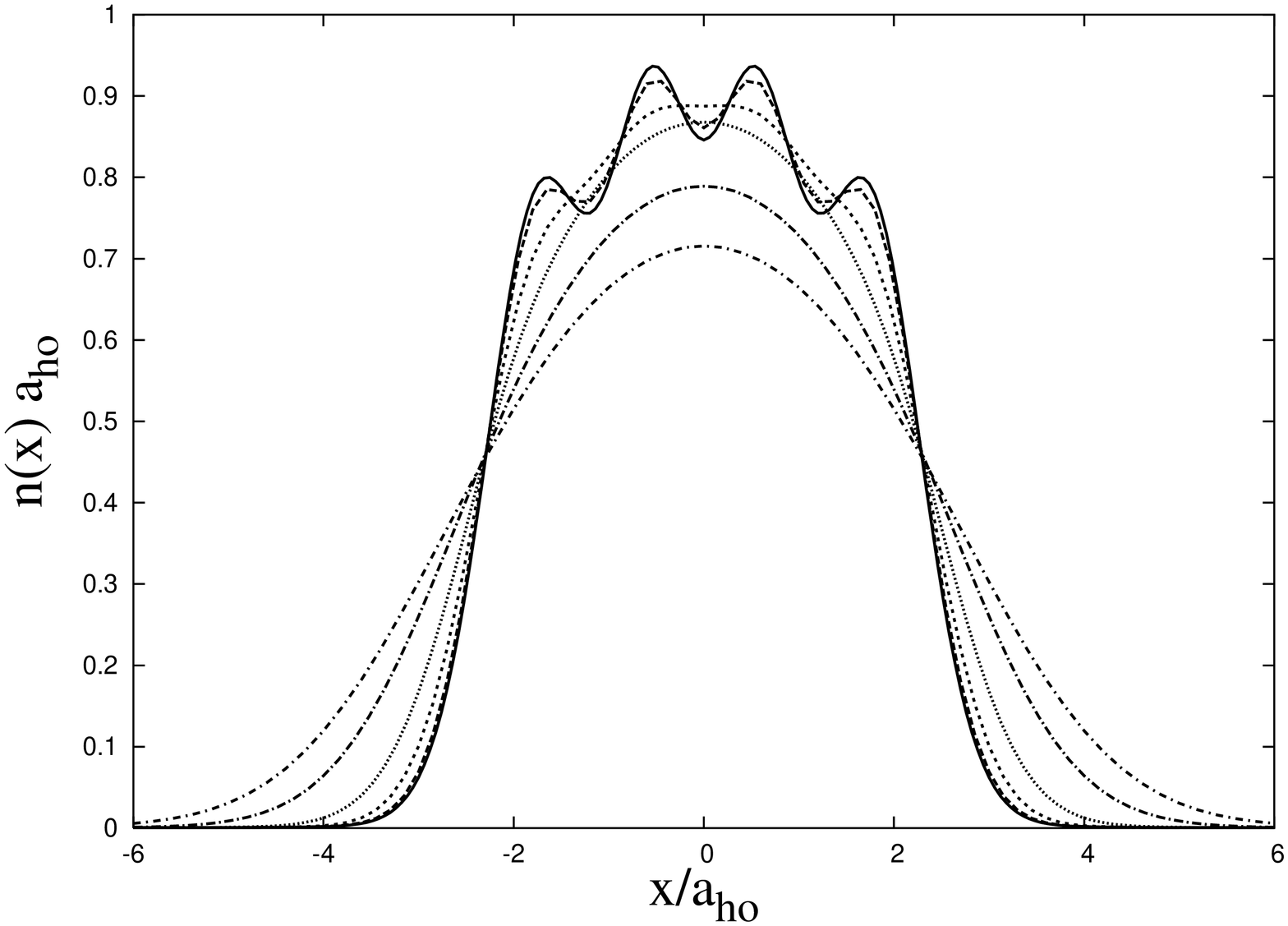,height=0.47\linewidth,angle=270}
\epsfig{file=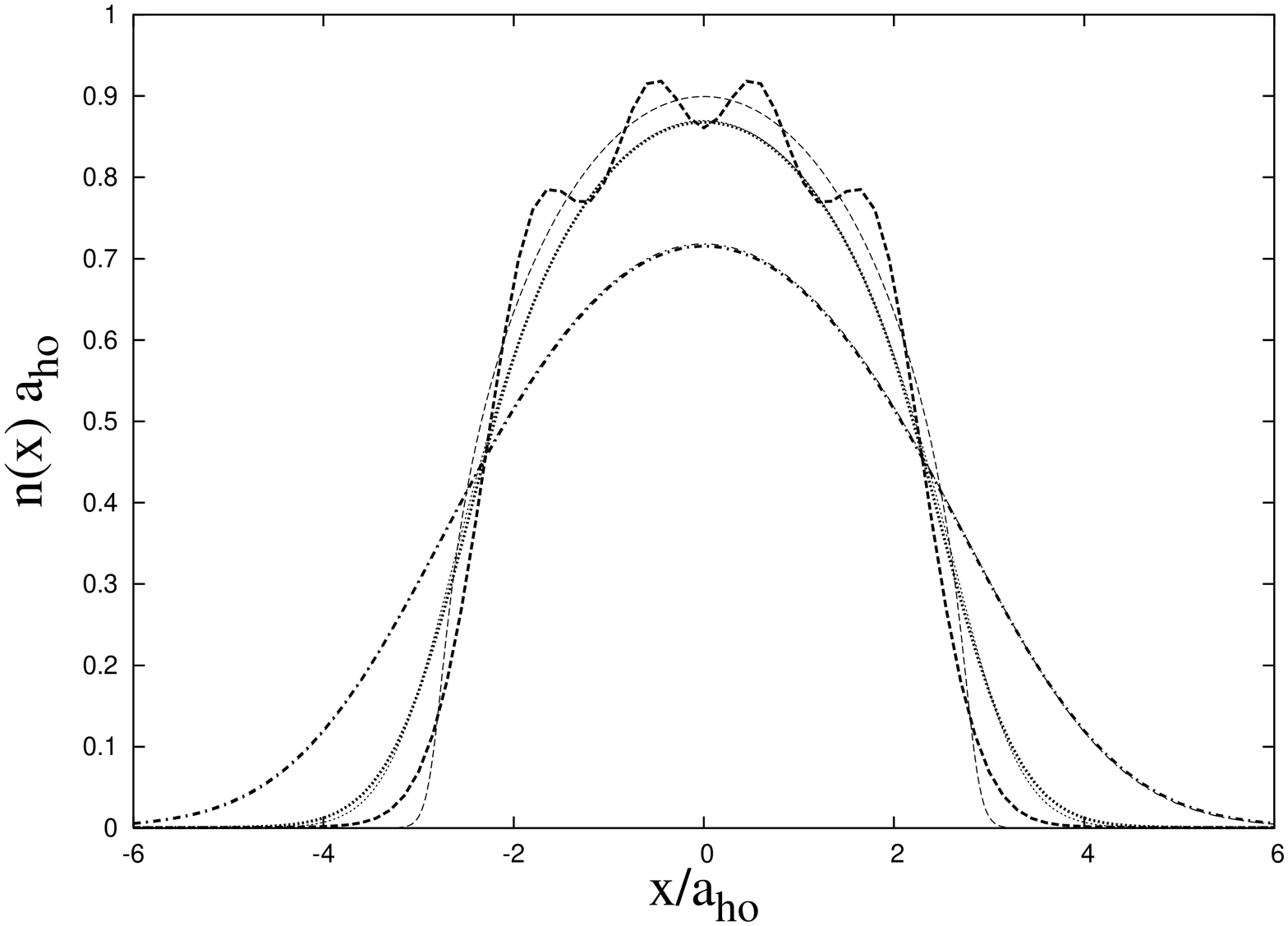,height=0.47\linewidth,angle=270}}
\vspace{1cm}
\caption{Particle density profile for 
$N=4$ harmonically confined fermions
at various values of the temperature. Left panel: ``exact'' profiles
at $T=0$ (solid curve, calculated by the method of Ref.
~\protect\cite{ref11}) and 
at $T=0.2\hbar\omega/k_B$ (dashed curve); the other curves
refer to $k_BT/\hbar\omega=0.5$, 1.0, 2.0 and 3.0, in order of decreasing peak
height. Positions are in units of the
 harmonic oscillator length 
$a_{ho}=\sqrt{\hbar/(m\omega)}$ and the particle density is in units of
$a^{-1}_{ho}$. The right-hand panel reports again the profiles at
$k_BT/\hbar\omega=0.2$, 1.0 and 3.0, together with those calculated in 
the semiclassical approximation (light lines); 
for the latter two values of $k_BT/\hbar\omega$ the ``exact'' profiles 
are hardly distinguishable from the semiclassical ones.}
\label{fig_rho_N4}
\end{figure}

\newpage
\begin{figure}
\centering{
\epsfig{file=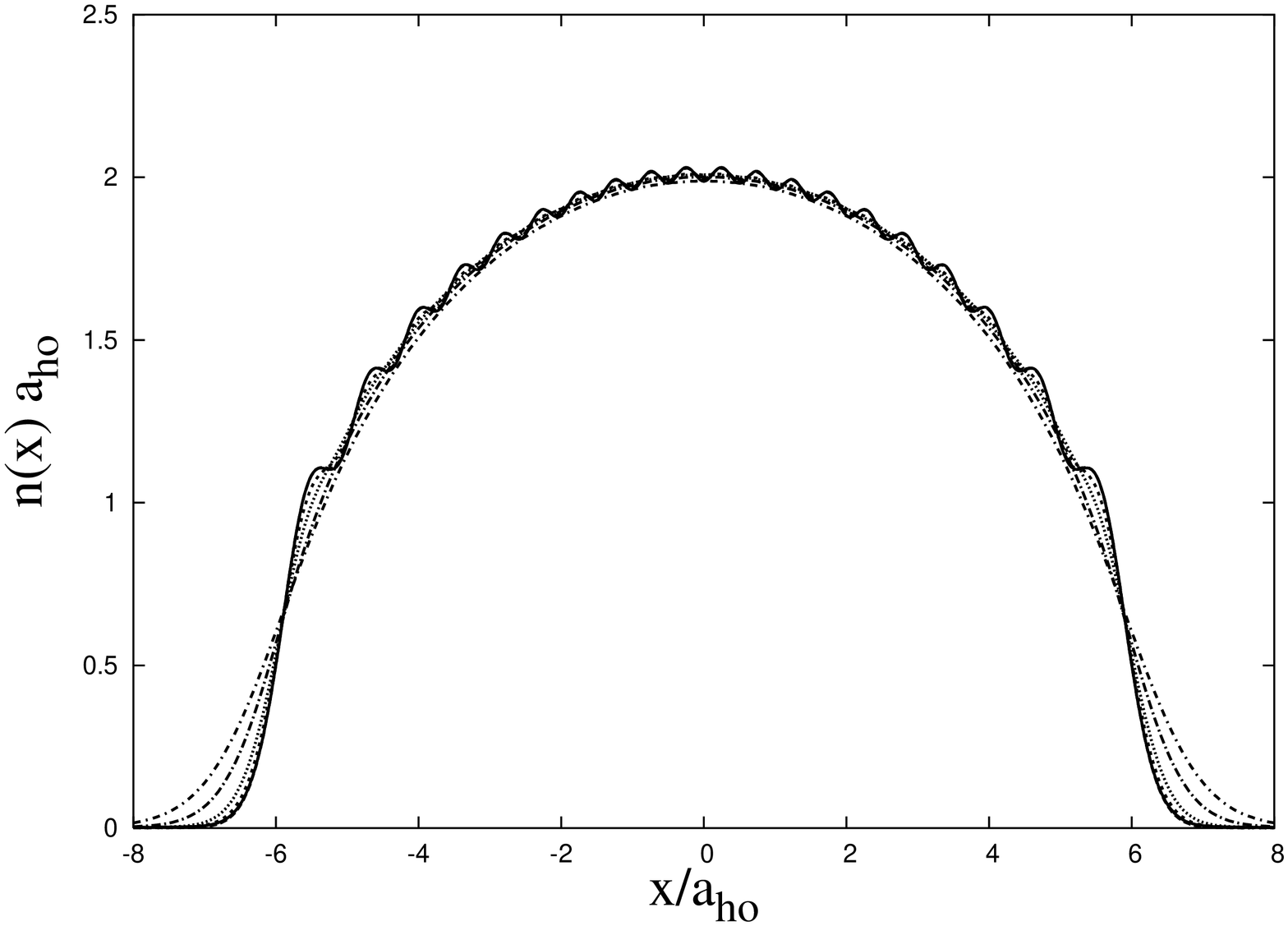,height=0.47\linewidth,
angle=270}
\epsfig{file=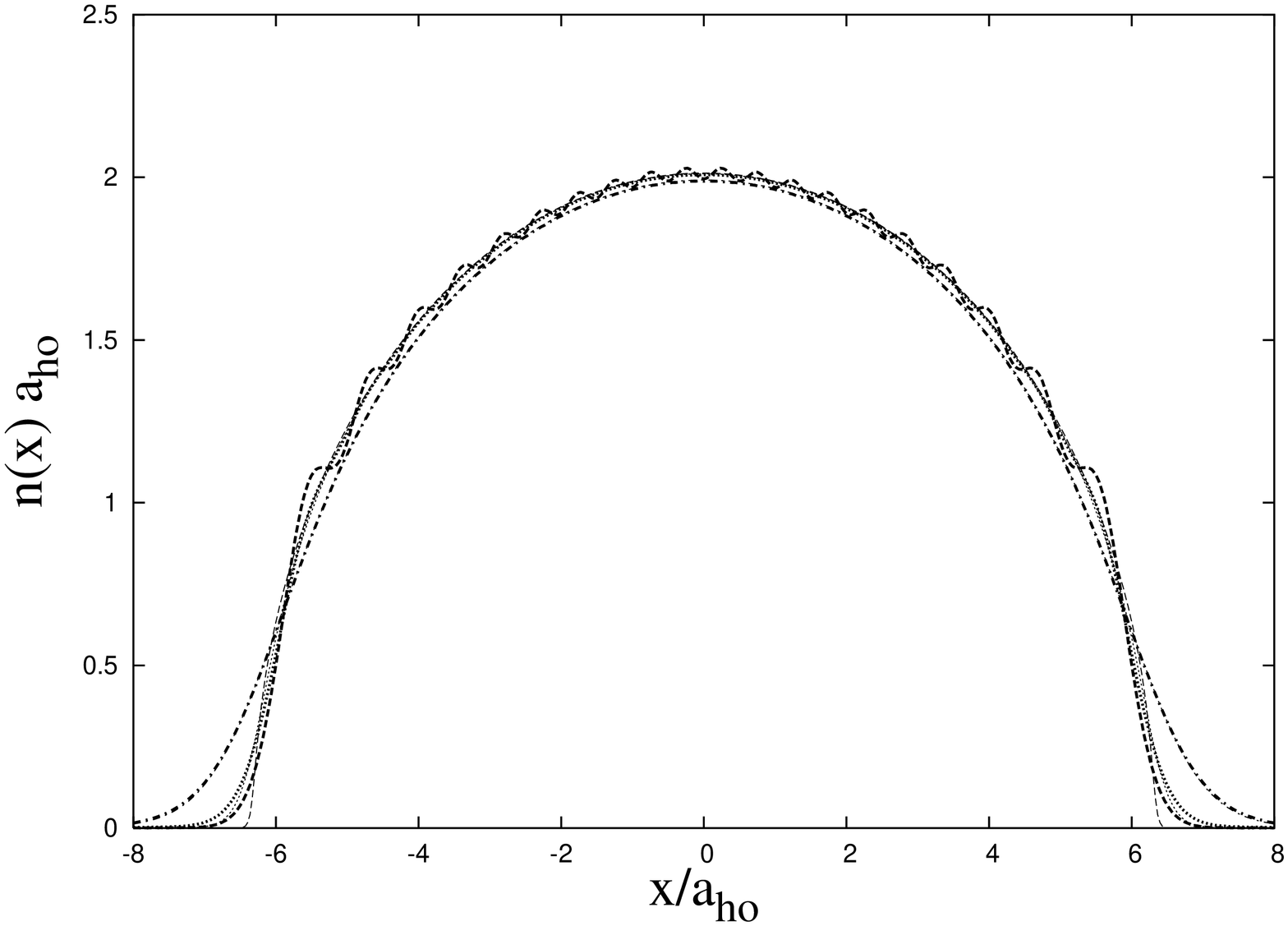,height=0.47\linewidth,
angle=270}}
\vspace{1cm}
\caption{The same as in Figure~\ref{fig_rho_N4}, for $N=20$ harmonically
confined fermions.}
\label{fig_rho_N20}
\end{figure}

\begin{figure}
\centering{
\epsfig{file=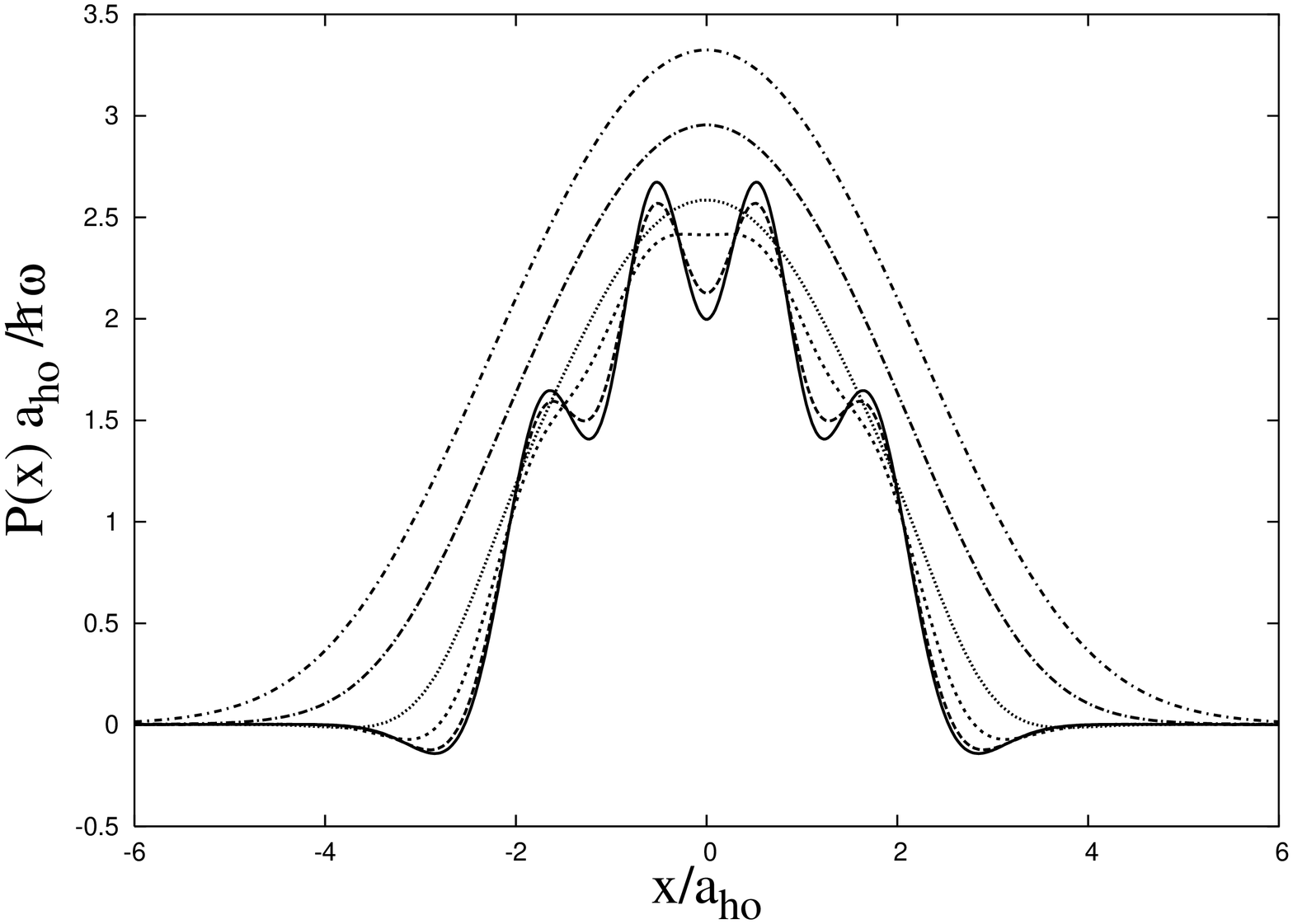,width=0.47\linewidth}
\epsfig{file=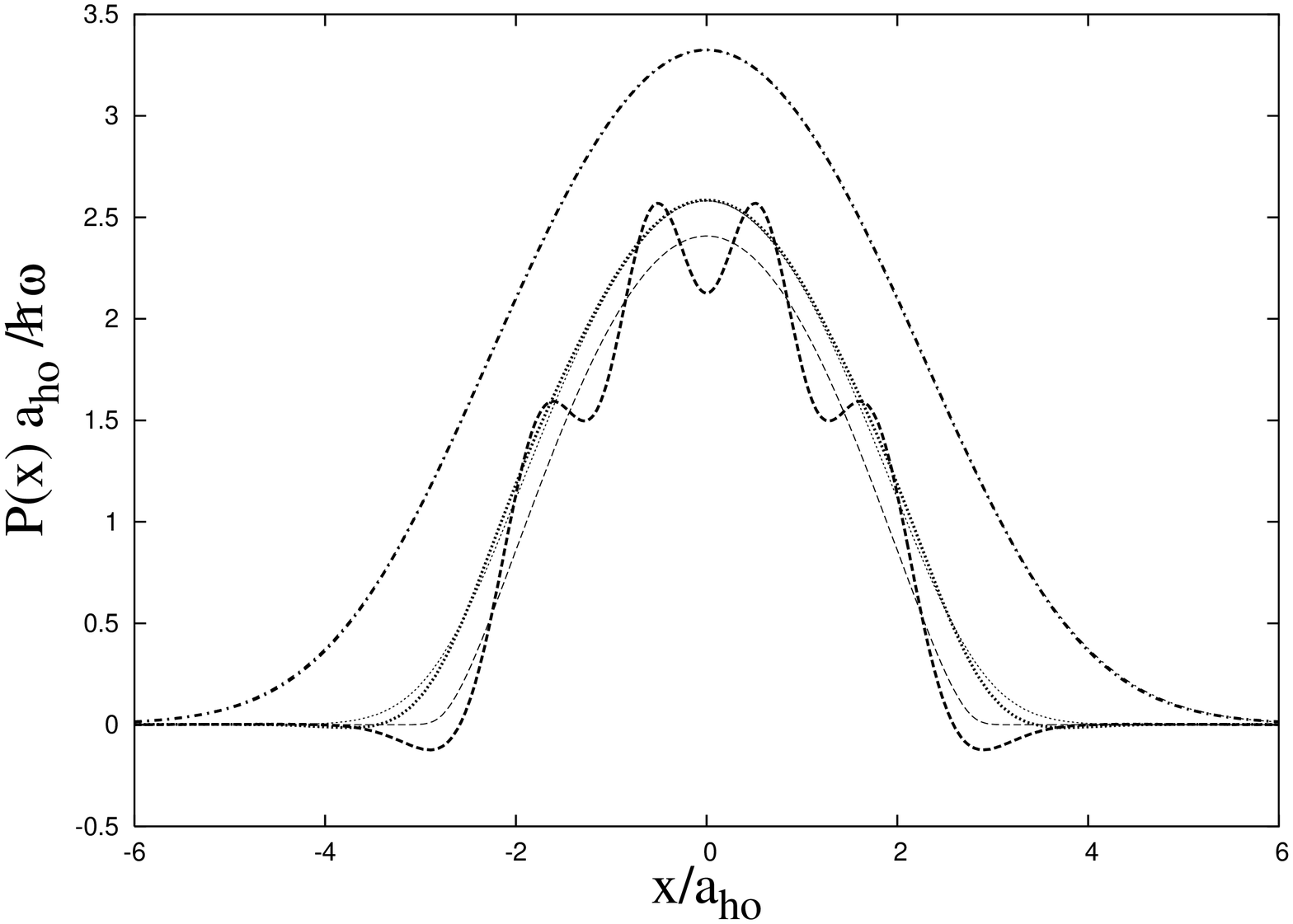,width=0.47\linewidth}}
\vspace{1cm}
\caption{Kinetic pressure profile for $N=4$ harmonically 
confined fermions
at the same values of the temperature as in Figure~\ref{fig_rho_N4}. Left panel:
``exact'' profiles; right panel: comparison with the semiclassical 
profiles for $k_BT/\hbar\omega=0.2$, 1.0 and 3.0
(light lines).
Positions are in units of 
$a_{ho}$ and the kinetic pressure is in units of
$\hbar\omega{a^{-1}_{ho}}$.} 
\label{fig_p2_N4}
\end{figure}

\newpage
\begin{figure}
\centering{
\epsfig{file=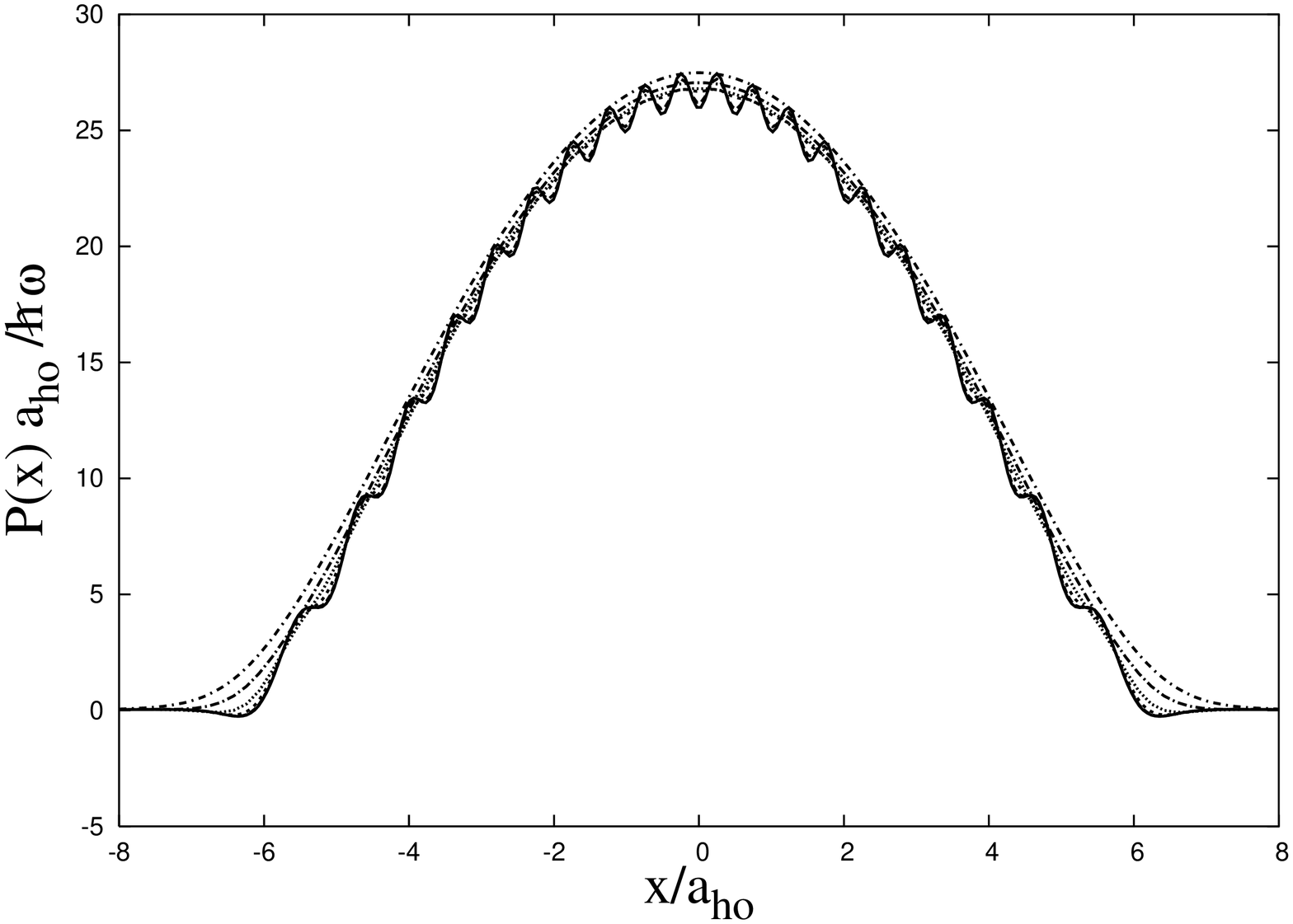,width=0.47\linewidth}
\epsfig{file=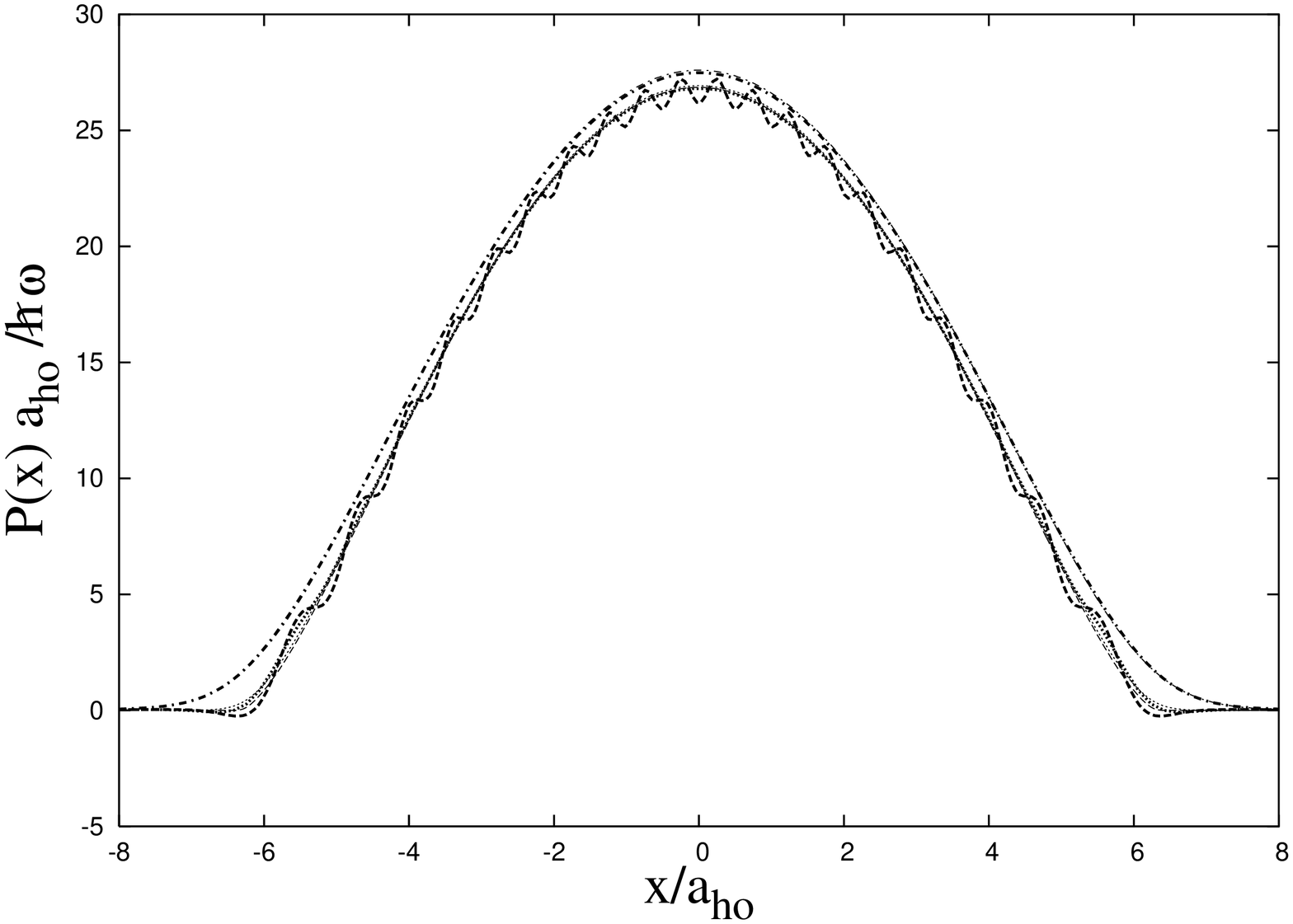,width=0.47\linewidth}}
\vspace{1cm}
\caption{The same as in Figure \ref{fig_p2_N4}, for $N=20$ harmonically 
confined fermions.}
\label{fig_p2_N20}
\end{figure}

\end{document}